\documentclass[a4paper,12pt]{article}
\usepackage[english]{babel}
\usepackage{graphicx}
\usepackage{color}
\usepackage[separate-uncertainty = true,multi-part-units=single]{siunitx}
\usepackage[margin=2cm]{geometry}
\usepackage{authblk}
\usepackage{amssymb}
\usepackage{amsmath}
\usepackage[numbers,sort&compress]{natbib}

\usepackage[labelfont=bf]{caption}

\begin{document}
	
	\title{Smectic and soap bubble optofluidic lasers}
	\date{}
	\author[1,2]{Zala Korenjak}
	\author[1,2,3]{Matja\v z Humar*}
	\affil[1]{Condensed Matter Department, J. Stefan Institute, Jamova 39, SI-1000 Ljubljana, Slovenia}
	\affil[2]{Faculty of Mathematics and Physics, University of Ljubljana, Jadranska 19, SI-1000, Ljubljana, Slovenia}
	\affil[3]{CENN Nanocenter, Jamova 39, SI-1000 Ljubljana, Slovenia}
	\affil[ ]{*E-mail: matjaz.humar@ijs.si}
	\affil[ ]{Keywords: soap bubble, whispering gallery modes, laser, liquid crystals, optofluidics}
	\maketitle

	\begin{abstract}
		
		Soap bubbles are simple, yet very unique and marvelous objects. They exhibit a number of interesting properties such as beautiful interference colors and the formation of minimal surfaces. Various optical phenomena have been studied in soap films and bubbles, but so far they were not employed as optical cavities. Here we demonstrate, that dye doped soap or smectic liquid crystal bubbles can support whispering gallery mode lasing, which is observed in the spectrum as hundreds of regularly spaced peaks, resembling a frequency comb. The lasing enabled the measurement of size changes as small as \SI{10}{nm} in a millimeter-sized, \SI{\sim 100}{nm} thick bubble. Bubble lasers were used as extremely sensitive electric field sensors with a smallest measurable electric field of \SI{110}{Vm^{-1}Hz^{-1/2}}. They also enable the measurement of pressures up to a \SI{100}{bar} with a resolution of \SI{1.5}{Pa}, resulting in a dynamic range of almost $10^7$. By connecting the bubble to a reservoir of air, almost arbitrarily low pressure changes can be measured while maintaining an outstanding dynamic range. The demonstrated soap bubble lasers are a very unique type of microcavities which are one of the best electric field and pressure microsensors to date and could in future also be employed to study thin films and cavity optomechanics.
		
	\end{abstract}
	
	\section*{Introduction}
	
	A soap bubble is made of a thin film composed of water and surfactants, which encloses air and forms a spherical shape \cite{isenberg1978science}. Bubbles are interesting from a variety of perspectives including mathematics, physics, chemistry and even biology, due to their similarity to the biological membranes. Soap bubbles were studied in terms of the interference colors \cite{afanasyev2011measuring}, geometry \cite{almgren1976geometry,taylor1976structure}, fluid motion \cite{debregeas1998life,meuel2013intensity}, mechanical oscillations \cite{kornek2010oscillations} and recently, branched flow of light \cite{patsyk2020observation}. A bubble can also be made purely from surfactant-like molecules. Specifically, smectic liquid crystals, which molecules form well defined molecular layers, are used for this purpose \cite{stannarius1998self,may2012dynamics,clark2017realization,lopez2011nematic}. The mixture of soap and water can actually also form liquid crystalline (lyotropic) phases. In fact, the origin of the word "smectic" is related to soap. Smectic bubbles have some very unique properties. Since the smectic liquid crystal molecules form well defined layers, the film thickness is quantized, that is to say, the film has an integer number of molecular layers. For example, bubbles with a completely uniform thickness of \SI{11}{nm} as large as \SI{1}{cm} in diameter were made \cite{stannarius1998self}. Therefore, quite uniquely, the ratio of the bubble size to its thickness can be as large as 6 orders of magnitude. The layered structure of the smectic bubbles makes their thickness completely stable and enables virtually infinite bubble lifetime as long as the air volume inside the bubble is kept constant. 
	
	Optical resonances called whispering gallery modes (WGMs) are formed when the light is trapped in a spherical object due to multiple total internal reflections and circulates near the surface of the sphere. WGMs were studied in various geometries including solid hollow cavities in the form of glass microbubbles \cite{sumetsky2010optical,watkins2011single} and glass capillaries \cite{shopova2007optofluidic}, which were employed for sensing applications \cite{ward2014hollow} and lasing \cite{shopova2007optofluidic,chen2016lasing}. However, WGMs were not studied until now in soap bubbles.

	Here we show that dye doped soap and smectic bubbles can support WGM lasing. Due to their fluid nature the bubbles are very soft compared to their glass counterparts, which influences the lasing and enables some unique applications.

	\section*{Results}
	
	\subsection*{Lasing of regular soap bubbles}

	Millimeter-sized soap bubbles doped with a fluorescent dye, inflated at the end of a capillary and illuminated with a pulsed laser were used to demonstrate the laser emission (Fig. \ref{fig1}a). To create the bubbles, a capillary was dipped into the soap solution and the air pressure in the capillary was briefly increased to inflate the bubble (Supplementary Video 1). The soap bubble could be later inflated or deflated to reach the desired size (\SIrange{0.4}{4}{mm}). Based on the interference colors \cite{afanasyev2011measuring} observed in reflection (Fig. \ref{fig1}b) the thickness of the soap film was typically in the range \SIrange{100}{800}{nm} and changed slowly in time as the bubble aged. The soap bubbles had a relatively uniform fluorescence intensity, except at the end of the capillary where there was a larger amount of the solution (Fig. \ref{fig1}c).

\begin{figure}[b!]
	\begin{center}
		\includegraphics[width=17cm]{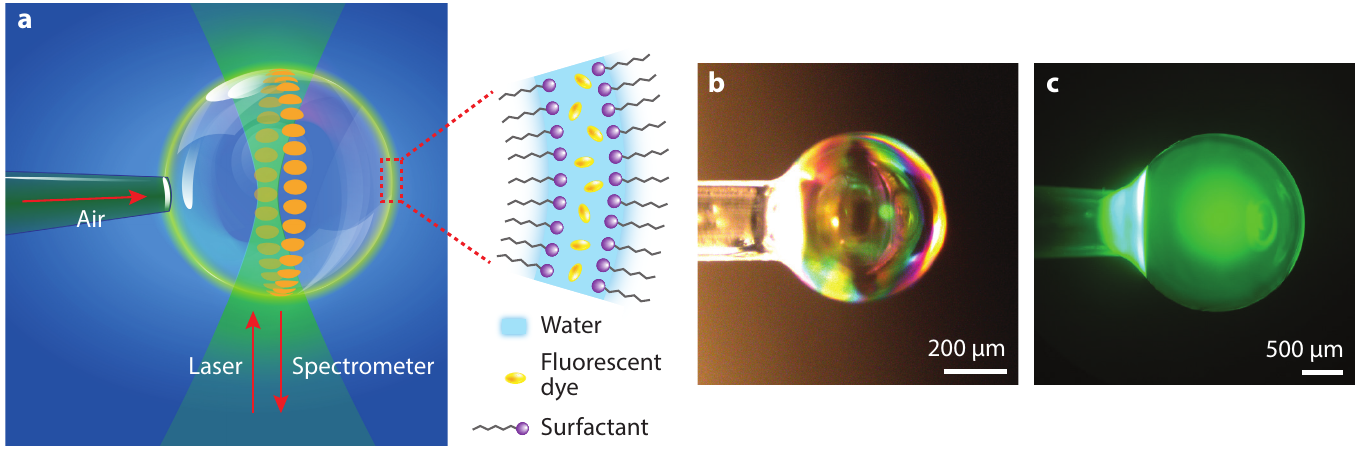}
		\caption{\textbf{A soap bubble formed at the end of a capillary.} (a) Scheme of the experimental configuration. A dye doped soap bubble is inflated at the end of a horizontal capillary and illuminated by a laser from below. The soap film is composed of a layer of water, surfactant molecules and fluorescent dye molecules. (b) A soap bubble in reflected light. Interference colors are visible. (c) Fluorescence image of a dye doped bubble.}
		\label{fig1}
	\end{center}
\end{figure}

	When a bubble was illuminated by a pulsed laser it emitted laser light due to the circulation of WGMs. When the center of the bubble was pumped, laser light was generated in all vertical planes except the ones intersecting with the capillary (Supplementary Fig. S1). This was observed as a bright rim, except on the opposite side of the capillary (Fig. \ref{fig2}a). When the bubble was illuminated at an edge, the generated light preferentially circulated in one plane, which was observed as a bright narrow ring (Fig. \ref{fig2}b).

	\begin{figure}[t!]
		\begin{center}
			\includegraphics[width=16cm]{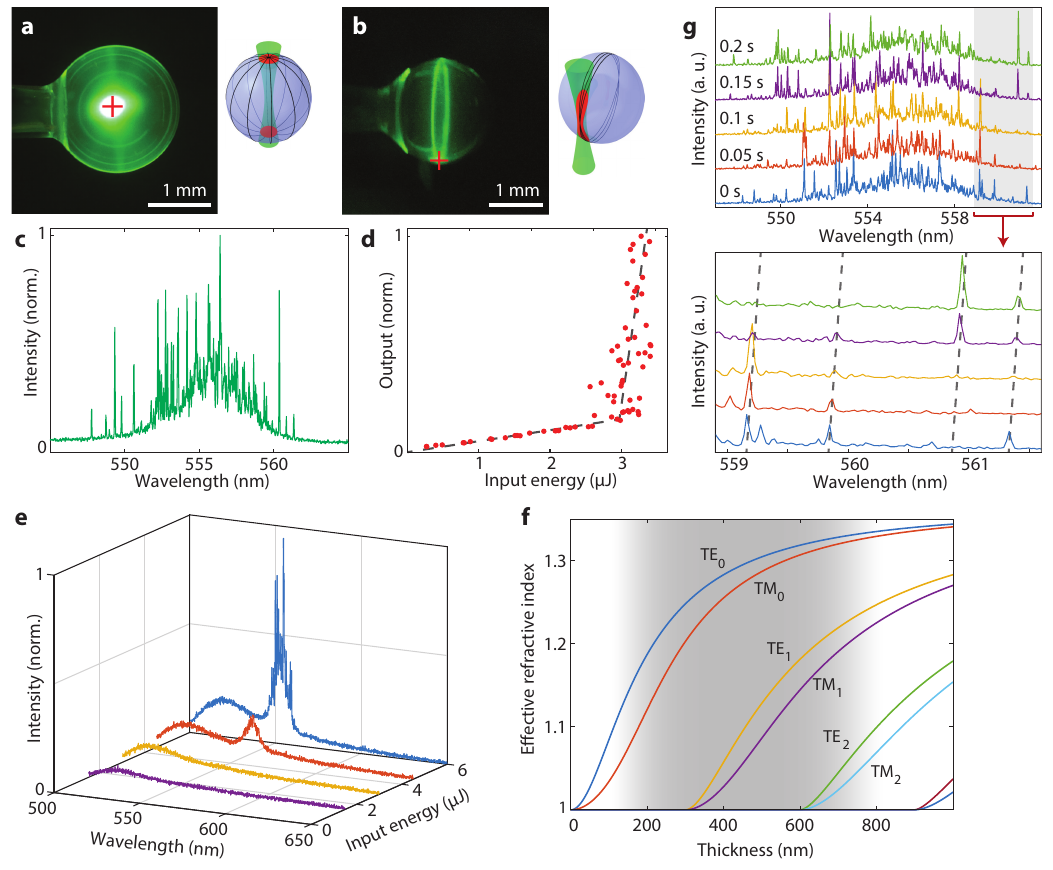}
			\caption{\textbf{Soap bubbles emitting laser light.} (a) A soap bubble illuminated in the center (red cross) by a pump laser. The drawing shows the gain regions (red) which are formed at the positions where the laser beam (green) passes through the soap film. The WGMs circulating in any vertical plane (black lines) pass through the two gain regions and are therefore excited. (b) Another bubble but illuminated at its rim, which generates a bright ring of circulating light. The gain region in this case is in the shape of a vertical patch. WGMs circulating in a narrow band in a vertical plane experience the most gain and are therefore observed as a bright ring. (c) A typical emission spectrum when a bubble is pumped above the lasing threshold. (d) Lasing intensity summed over a \SI{5}{nm} wide spectrum range as the input laser pulse energy was increased shows a typical threshold behavior. The dashed lines are a guide to the eye. (e) Spectrum emitted by a soap bubble attached to the end of a capillary when the pulse energy of the pump laser is increased. At lower energies only fluorescence is observed, while at higher energies sharp peaks appear. (f) Slab waveguide modes for a soap film at the wavelength of \SI{555}{nm}. The shaded area corresponds to the approximate thickness range of the bubbles used in the experiments. (g) Spectra of consecutive pump laser pulses. Displaying a smaller wavelength range reveals a shift of the modes towards longer wavelengths.}
			\label{fig2}
		\end{center}
	\end{figure}

	Sharp spectral lines were present in the spectrum of the emitted light (Fig. \ref{fig2}c). A clear lasing threshold was observed when the pump laser energy was increased (Fig. \ref{fig2}d and e). Typical thresholds were in the order of a few \SI{}{\micro J}. The WGMs can be approximated as
	\begin{equation}
	2\pi r n_{\mathrm{eff}}= \lambda l,
	\label{n_eff}
	\end{equation}
	where $r$ is the radius of the bubble, $n_{\mathrm{eff}}$ is the effective refractive index of the mode, $\lambda$ is the wavelength of the mode and $l$ is the azimuthal mode number. Since the bubbles were usually relatively large (\SI{\sim1}{mm}) compared to the wavelength of the light, as a first approximation, the light propagation can be described as a flat slab waveguide. The effective refractive indices (Fig. \ref{fig2}f) were calculated using standard equations for light propagation in a flat slab waveguide \cite{kawano2004introduction}, where the slab is the soap film ($n=1.364$) and both sides are air ($n=1$). Calculating the positions of WGMs by using the effective index for flat slab is accurate to a few percent compared to the WGMs simulations \cite{zhang2017chip}. For a typical bubble with a thickness of \SI{600}{nm} the effective refractive indices were 1.319 and 1.307 for first radial order TE and TM modes, respectively. For a typical soap bubble thickness not only the fundamental, but also higher guided modes may be present. In the lasing spectrum no regularly spaced peaks could be observed. This is probably due to a large number of possible modes and irregularities of the soap film. With each pulse of the pump laser the spectrum changed slightly (Fig. \ref{fig2}e). Some modes disappeared, while some others appeared, but there was a general trend towards a shift towards longer wavelengths. The shifting of the peaks in time was probably caused by a combination of size, thickness and refractive index changes. However, from the spectrum alone we can not determine how each of the three parameters contributes to the shift in the spectrum.

 \begin{figure}[htb]
		\begin{center}
			\includegraphics[width=16cm]{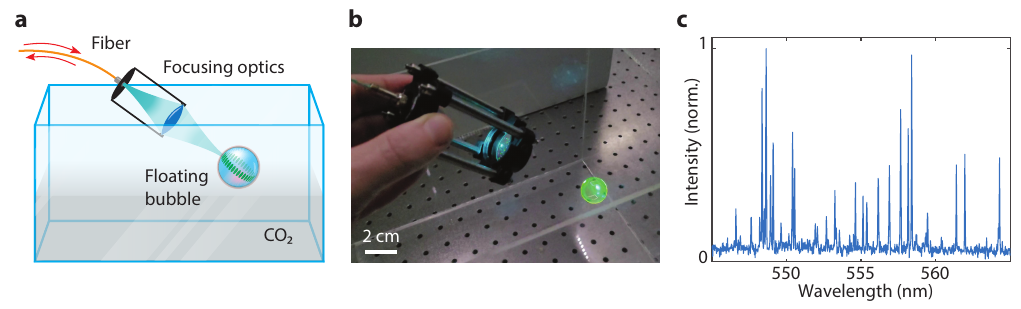}
			\caption{\textbf{Free floating bubbles.} (a) Experimental setup for the free floating soap bubbles. (b) A photo of a larger floating soap bubble, which is illuminated by the pump laser, and emits laser light. (c) Spectrum from a \SI{\sim2}{cm} diameter free floating bubble.}
			\label{fig3}
		\end{center}
	\end{figure}
	Lasing was also achieved in free floating soap bubbles. A bubble was blown into a tank filled with CO$_2$ so that it was floating on the surface of the CO$_2$. The bubble was illuminated by the pump laser by using an optical fiber and a lens (Fig. \ref{fig3}a and b). The same lens and optical fiber were also used to collect the generated light and send it to the spectrometer. Spectra with sharp spectral lines corresponding to lasing were also observed in this case (Fig. \ref{fig3}c).

	\subsection*{Lasing of smectic bubbles}

	To better control the thickness and the refractive index of the film, bubbles made out of smectic liquid crystal were employed. Specifically, 8CB was used, which has a smectic A phase at room temperature. Under transmitted light the smectic bubbles looked completely uniform (Fig. \ref{fig4}a) indicating that the whole surface had thickness of equal number of molecular layers (Fig. \ref{fig4}b). If some region had one more or one less molecular layer this would result in a well visible island or hole, respectively \cite{clark2017realization}. The thickness of the bubbles could be approximately controlled by how fast they were inflated, with the faster inflation resulting in a thinner bubble. Typically the bubbles employed in this study had a thickness in the range of \SIrange{30}{120}{nm}, which was determined from the intensity of the transmitted light \cite{stannarius1999self}.

	Imaging a bubble in between crossed polarizers \cite{mirri2015situ} indicates an uniform radial orientation of the molecules (Fig. \ref{fig4}c and d).  
	
	\begin{figure}[t!]
		\begin{center}
			\includegraphics[width=9cm]{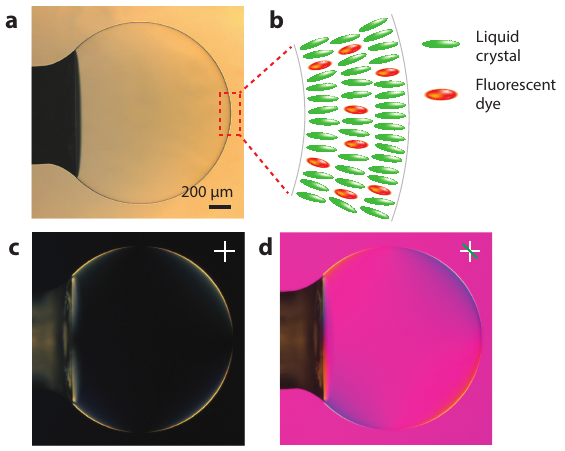}
			\caption{\textbf{Smectic bubbles.} (a) A smectic bubble in transmitted light. (b) Scheme of the molecular structure of the film. Drawn is an example of a three layer smectic film composed of ordered liquid crystal and dye molecules. (c) Same bubble under crossed polarizers. The typical bright cross is observed indicating uniform molecular orientation either parallel or perpendicular to the bubble surface. (d) Same bubble with an additional waveplate inserted between the polarizers. From the resulting colors it can be deduced that the molecules are oriented perpendicular to the surface, that is in the bubble radial direction. 
			}
			\label{fig4}
		\end{center}
	\end{figure}

	\begin{figure}[h]
	\begin{center}
		\includegraphics[width=10cm]{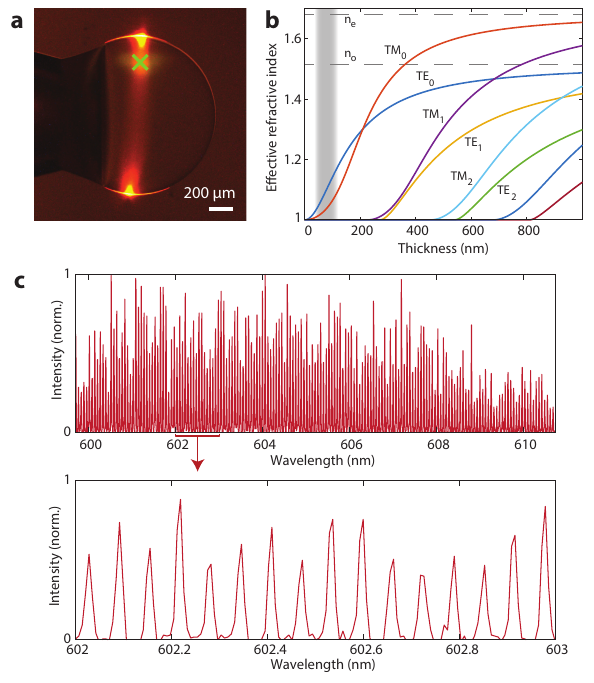}
		\caption{\textbf{Lasing of smectic bubbles.} (a) Image of a lasing smectic bubble. The cross indicates the pump laser beam position. (b) Effective refractive indices for different modes calculated for a slab waveguide at the wavelength of \SI{610}{nm}. The shaded area correspond to the approximate thickness range of the bubbles used in the experiments. The bulk refractive indices of TE and TM modes are different (dashed lines). (c) Frequency comb-like spectrum of a lasing smectic bubble \SI{1.75}{mm} in diameter.}
		\label{fig5}
	\end{center}
\end{figure}

\begin{figure}[htb]
	\begin{center}
		\includegraphics[width=9cm]{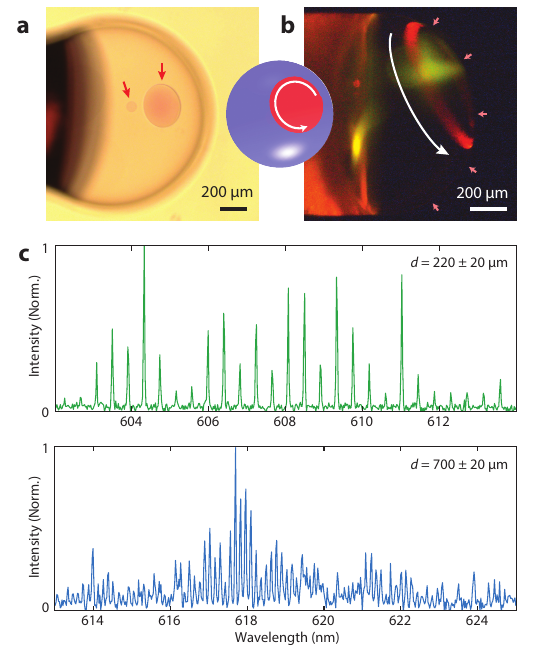}
		\caption{\textbf{Lasing of smectic islands.}  (a) Two smectic islands (indicated by the arrows) on a bubble. (b) Image of a lasing island observed as a red ring (white arrow) on the surface of the bubble (indicated by the red arrows). (c) Lasing spectra of two smectic islands of a different diameter $d$.}
		\label{fig6}
	\end{center}
\end{figure}

	When pumping with a pulsed laser, the dye doped smectic bubbles emitted laser light visible as a ring and two bright spots (Fig. \ref{fig5}a). The smectic bubbles were extremely stable and the experiments on a single smectic bubble could be performed for up to \SI{30}{min}. For a typical bubble thickness (\SIrange{30}{120}{nm}) only TM$_0$ and TE$_0$ modes exist (Fig. \ref{fig5}b). Since the smectic film is birefringent ($n_\mathrm{o}=1.51$ and $n_\mathrm{e}=1.68$), the bulk refractive indices of TE and TM modes are different. The effective refractive indices were calculated in the same way as for soap bubbles, by using standard equations for light propagation in a flat slab waveguide \cite{kawano2004introduction}, but using the following refractive indices: $n_\mathrm{air}=1$ on both sides of the film and for the smectic film, $n_\mathrm{o}$ for TE polarized light and $n_\mathrm{e}$ for TM polarized light. At a thickness of \SI{50}{nm} the TE$_0$ has a significantly larger effective refractive index (1.049) than TM$_0$ mode (1.014) and is the only mode lasing, as also identified by measuring the emission intensity through a polarizer.
	
	At this typical bubble thicknesses higher radial modes are not allowed. This is a significant advantage over bulk WGM microcavities which will typically display lasing of a number of different modes when their size is large. On the contrary, the emission spectrum above the lasing threshold was very regular with equally spaced spectral lines, resembling a frequency comb (Fig. \ref{fig5}c). The emission typically spanned \SIrange{10}{20}{nm}. For a \SI{1.8}{mm} diameter bubble the average free spectral range (FSR) was \SI{0.06}{nm} and the spectrum contained $\SI{\sim 250}{}$ lasing peaks. The azimuthal mode number of WGMs at the wavelength of \SI{605}{nm} is $l\approx 10\,000\pm100$. The contrast between the fluorescent background and the lasing peaks in the spectrum was excellent with the background being practically zero. From the measured FSR and the diameter of the bubble determined from the image (\SI{1.90 \pm 0.02}{mm}), the effective refractive index is calculated to be $1.06 \pm 0.01$. For TE modes, using equations for a slab waveguide, this corresponds to a bubble thickness of \SI{57 \pm 5}{nm}. Smectic bubbles thinner than \SI{\sim 30}{nm} did not generate laser emission due to too low effective refractive index causing significant radiative light leakage and the fact that a thin layer also results in very few dye molecules to be present, resulting in a low gain. The estimated gain length of Pyrromethene 597 recalculated for a \SI{100}{nm} thick slab waveguide is \SI{180}{\micro m} \cite{vellaichamy2023optical,lyu2021correction} which is much larger compared to the bubble thickness.

	The size of smectic bubbles can be changed in real time, within a few seconds, by inflating or deflating them in at least the size range \SIrange{0.5}{5}{mm} resulting in a large tunability of the FSR in the range of \SIrange{0.02}{0.2}{nm}. This enables to use them as largely tunable frequency-like light source. This tunability is significantly larger than for their solid state counterparts.

	When the volume of the bubble is increased or decreased very quickly, stable circular floating islands with larger thickness than the rest of the bubble can be formed (Fig. \ref{fig6}a) \cite{clark2017realization,stannarius1998self}. The islands can also be created by illuminating a thicker bubble ($\gtrsim\SI{200}{nm}$) with a pulsed laser, where each pulse can create one island (Supplementary Fig. S3). The size of the islands can be from a few micrometers up to being as large as the bubble itself. Islands larger than \SI{\sim 100}{\micro m} emitted laser light upon excitation with a pulsed laser by the circulation of light around the perimeter of the island (Fig. \ref{fig6}b). The resulting WGM lasing was observed as periodic peaks in the spectrum (Fig. \ref{fig6}c). A typical effective refractive index calculated for the island size and FSR was $1.3 \pm 0.1$. This enabled stable laser emission with periodic lasing peaks with a wide range of FSR values (\SIrange{0.1}{0.9}{nm}) and azimuthal mode numbers from 700 to 6100 depending on the size of the islands.
	
	\newpage
	
	\subsection*{Precise measurement of the smectic bubble laser size}

\begin{figure}[t]
		\begin{center}
			\includegraphics[width=14cm]{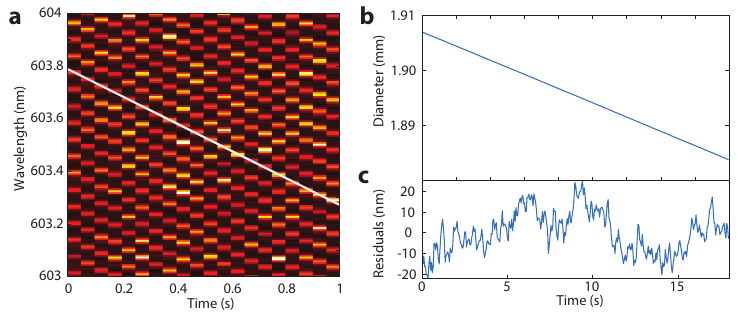}
			\caption{\textbf{Measurement of smectic bubble size.} (a) Lasing spectrum from a \SI{1.75}{mm} smectic bubble in time. The color represents the intensity. The white line highlights the shifting of a single lasing mode in time due to the decreasing size of the bubble. (b) Change in the diameter of a bubble in time measured by following the lasing peaks in a. The initial size was measured from the FSR. The bubble size was slowly decreasing due to air diffusion through the thin wall. Since the size measurement is very precise, the plot looks like a perfectly smooth line. (c) When the decrease in size is subtracted from the data, only an extremely small noise remains.}
			\label{fig7}
		\end{center}
	\end{figure}

	The changes in the bubble size were measured by following shifts of the lasing peaks (Fig. \ref{fig7}a). The relative size change is approximately equal to the relative wavelength shift: 
	\begin{equation}
		\Delta d/d \approx \Delta \lambda/\lambda.
		\label{d_lambda}
	\end{equation}
	From the size measurement in time it is visible that the size of the bubble slowly decreases in time (Fig. \ref{fig7}b) due to the diffusion of air molecules through the thin wall (APPENDIX A) \cite{ishii2009gas}. For real applications the size decrease could be compensated by pumping additional air into the bubble, stabilizing the size, or simply reinflating the bubble when it becomes too small, thus enabling almost indefinitely long measurements. The diameter change rate was \SI{1.1}{\micro m/s} for a \SI{1.9}{mm} bubble. The time dependence was fitted to the equation for the air diffusion Eq. \eqref{R(t)_approx} and subtracted from the data. The remaining residuals are extremely small with a standard deviation of only \SI{10}{nm} (Fig. \ref{fig7}c). This means that size changes as small as \SI{10}{nm} can be measured, corresponding to an exceptional relative accuracy of $5 \times 10^{-6}$. A part of the observed noise may come from the environment, such as air pressure fluctuations and air currents, so that the actual optically limited measurement noise may be even lower than observed. To be able to track the spectral shifts, the shift in between two measurements (laser pump pulses) should be smaller than half the FSR. In the case of faster changes, FSR was measured instead, which enabled the measurement of arbitrarily large and fast size changes with a still excellent accuracy of \SI{50}{nm}. By taking into account the typical effective refractive index, the absolute size of the bubble can be measured via the FSR value by using Eq. \eqref{n_eff}.

	\subsection*{Smectic bubble lasers as electric field sensors}
	
	The smectic bubble lasers are very soft and therefore sensitive to external factors, which can change their size and shape. Simultaneously these changes can be measured very precisely via the lasing spectrum. Therefore, this makes soap bubble lasers excellent sensors.
	
	\begin{figure}[b!]
		\begin{center}
			\includegraphics[width=7cm]{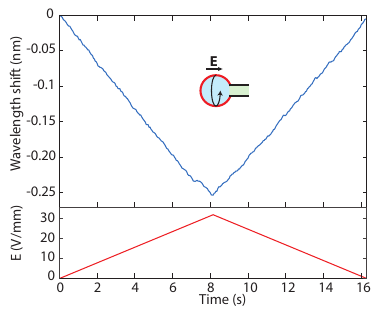}
			\caption{\textbf{Electric field measurement with smectic bubble lasers.} Shift of the lasing modes when an electric field is increased continuously to \SI{30}{V/mm} and back to zero. The continuous decrease in the size due to diffusion was subtracted from the data.}
			\label{fig8}
		\end{center}
	\end{figure}

	The bubble lasers were employed as electric field sensors and electrically tunable laser sources. Already applying less than \SI{1}{V} across the bubble deformed it \cite{ishii2011anomalous} and caused a measurable shift in the laser peaks (Fig. \ref{fig8}). The sensitivity was \SI{0.008}{nm(V/mm)^{-1}}, which is on par or larger than best previous reports using WGM microcavities \cite{kiraz2008spectral,ali2012high,humar2009electrically}. The minimum measurable field determined from the measurement noise and the sensitivity was \SI{110}{Vm^{-1}Hz^{-1/2}}. With this, our sensors outperforms most micro DC electric field sensors including sensors based on microelectromechanical systems \cite{kainz2018distortion,han2022micro}, piezoelectric \cite{xue2019piezoelectric}, optical \cite{zhang2021review,calero2019ultra,zhu2015high} and nitrogen-vacancy centers in diamond \cite{dolde2011electric,chen2017high}. Further, in contrast to some other electric field sensors, the bubbles are not conducting and therefore do not distort the measured electric field.

	\subsection*{Smectic bubble lasers as pressure sensors}
	
	The compressibility of the air within the bubble and the ability to measure the size extremely precisely enables the use of the bubbles as pressure sensors. The spectral shift for pressure change is $\Delta \lambda/\lambda \approx - \Delta p/ 3 p_0$, where $\Delta p$ is the pressure change and $p_0$ is the initial pressure. In all the following measurements and calculations the initial pressure was equal to the atmospheric pressure. This gives a size independent sensitivity $\Delta \lambda/\Delta p$ of \SI{2e-3}{nm/Pa}. This is several orders of magnitude more than previously reported for WGM microcavities, such as hollow polymer spheres (\SI{2e-6}{nm/Pa}) \cite{ioppolo2007pressure} and a glass microbubbles (\SI{4e-7}{nm/Pa}) \cite{henze2011tuning}.

\begin{figure}[htb]
		\begin{center}
			\includegraphics[width=17cm]{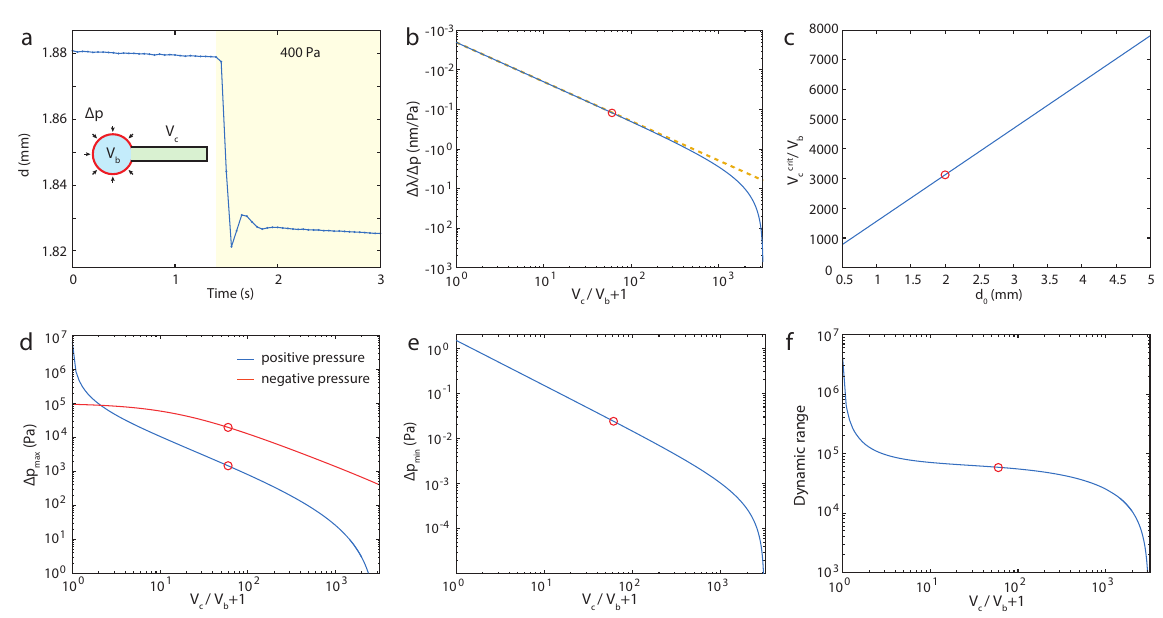}
			\caption{\textbf{Pressure measurement with smectic bubble lasers.} (a) Change of the bubble size measured from the lasing spectrum as the pressure outside the bubble was increased by \SI{400}{Pa} above the atmospheric pressure. The volume of the capillary in this experiment was \SI{80}{\micro l} and the initial volume of the bubble was \SI{4.2}{\micro l}, resulting in $V_c/V_b+1\approx20$ for this particular case. (b) The calculated pressure sensitivity of a \SI{2}{mm} bubble as a function of total volume relative to the volume of the bubble ($V_c/V_b+1$). The dotted line would be for a bubble with zero surface tension. The red circle in all panels represents the maximum $V_c$ used in the experiments. (c) Critical additional volume relative to the volume of the bubble, at which the bubble becomes unstable. Above this additional volume the bubble collapses due to the Laplace pressure. (d) Maximum measurable positive and negative pressure. (e) The minimum measurable pressure change. (f) The dynamic range for positive pressure change, that is the ratio between the highest and the lowest measurable pressure changes.}
			\label{fig9}
		\end{center}
	\end{figure}

	To test the performance of the pressure sensor, the pressure around the bubble was controllably changed and the diameter of the bubble was calculated from the measured FSR value. When the pressure was increased by \SI{400}{Pa} above the atmospheric pressure the bubble diameter decreased by \SI{52}{\micro m} (Fig. \ref{fig9}a). In all experiments the bubble was connected to a capillary which provided an additional fixed volume of air. Since a larger volume of air is easier to compress, this significantly increases the pressure sensitivity.

	A pressure difference between the interior and the exterior of the bubble is given by the Laplace law. For a small change in the external pressure, the size of the bubble changes such that the pressure difference satisfies Laplace condition.
	The pressure in the bubble is changed due to the change in the volume of the bubble.
	If we assume that the process is isothermal, the pressure inside the bubble after the change, $p_i'$ is:
	\begin{equation}
	\frac{p_i'}{p_i}=\frac{V}{V'},
	\label{isothermal}
	\end{equation}
	where $p_i$ is the initial pressure in the bubble, $V$ and $V'$ are the initial and final volumes of the bubble, respectively.	Since the bubbles are connected to the capillary, the volume of both the capillary ($V_c$, which remains unchanged) and the bubble ($V_b$) must be included in Eq. \eqref{isothermal}, therefore $V=V_c+V_b$ and $V'=V_c+V_b'$.
	The Laplace pressure can be written explicitly as the difference between the pressures on both sides of the film:
	\begin{align}
	\label{p_L}
	p_L&=\frac{4\sigma}{R}=p_i-p_0,\\ 
	\label{p_L'}
	p_L'&=\frac{4\sigma}{R'}=p_i'-(p_0+\Delta p),
	\end{align}
	where $p_L$ and $p_L'$ are Laplace pressures before and after the change, $p_0$ is the initial external pressure, and $\Delta p$ is the change in the external pressure.
	Combining  Eq. \eqref{isothermal}, \eqref{p_L} and \eqref{p_L'} an equation that relates the change in external pressure to the change in bubble size is obtained:
	\begin{equation}
	\Delta p = \frac{V}{V'}p_L - p_L' + p_0\left(\frac{V}{V'}-1\right). \label{dp}
	\end{equation}
	In the approximation of small radius change, the relation between the small size and pressure change is
	\begin{equation}
	\frac{\Delta R}{\Delta p} = - R\left[\frac{3}{V_c/V_b+1}\,\left(p_0+\frac{4\sigma}{R}\right) - \frac{4\sigma}{R}\right]^{-1}.
	\end{equation}
	
    The volume of the bubble is also changing because of the diffusion of air molecules through the thin wall. Therefore if the measured pressure change is not significantly faster than the diffusion, this needs to be taken into account. Specifically, the shrinkage rate depends on the size and the size is dependent on the pressure and diffusion. To account for both contributions, Eq. (\ref{dp}) and (\ref{R(t)}) should be combined. It is worth noting that the sensitivity for $V_c=0$ is not dependent on the size of the bubble, so not dependent on the air diffusion. On the other hand, for $V_c/V_b+1=60$ the sensitivity changes for $12\%$ in one minute, which is significant and should be taken into account.

    Further, the temperature change of the air has the same effect on the bubble volume as the pressure change, so it needs to be taken into account as well (APPENDIX B). In practice, a separate thermometer with a resolution of at least \SI{6}{mK} could be used to compensate for temperature changes.

    The sensitivity is calculated by solving Eq. \eqref{dp} while taking into account Eq. \eqref{d_lambda}. At small additional volumes the sensitivity is inversely proportional to the total volume of the system  $V_c+V_b$, or more conveniently written as the total volume relative to the volume of the bubble $V_c/V_b+1$ (Fig. \ref{fig9}b). In the experiments up to $V_c/V_b+1=60$ was used providing a sensitivity of \SI{0.12}{nm/Pa}.
	
	Above approximately $V_c/V_b+1=1000$, for a surface tension of \SI{0.024}{N/m}, the surface tension starts to have a significant contribution and make the bubble unstable. At $V_c$ higher than $V_c^{crit}$ the bubble is not stable any more and collapses by itself due to the Laplace pressure. $V_c^{crit}$ is dependent on the surface tension and size of the bubble (Fig. \ref{fig9}c). At the critical volume the sensitivity diverges. Therefore by having a very large additional volume, an arbitrarily high sensitivity can be in principle achieved. In the experimental case ($V_c/V_b+1=60$) the critical volume is $V_c^{crit}/V_b+1=3130$.
	
	The bubble also collapses at any $V_c>0$ if a large enough positive outside pressure is applied. This maximum measurable pressure change (Fig. \ref{fig9}d) depends on the additional volume in the capillary $V_c$ and is calculated by solving Eq. \eqref{dp}. It is the maximum pressure change at certain $V_c$ at which the solutions exist. The maximum measurable positive pressure change is also limited by the minimum size of the bubble which is still capable of lasing (\SI{\sim 0.5}{mm}). The maximum measurable negative pressure change is limited by the maximum size of the bubble (\SI{\sim 5}{mm}) for which the spectral lines can still be distinguished by the spectrometer.

	The resolution of the measurements is the smallest detectable pressure change (Fig. \ref{fig9}e) and was calculated from the sensitivity and the smallest measurable wavelength shift. By increasing $V_c$ the smallest detectable pressure change can be almost arbitrarily small, however, at the expense of the maximum measurable pressure.  For a stand alone bubble ($V_c=0$) the smallest measurable pressure change is \SI{1.5}{Pa}, while the largest positive pressure change is almost \SI{e7}{Pa}, that is \SI{100}{bar}. Namely, a pressure change of \SI{e7}{Pa} would shrink a bubble from \SI{3}{mm} to \SI{0.65}{mm}, well within the lasing regime. This gives an exceptionally large dynamic range of almost $10^7$. For the largest additional volume used in the experiments ($V_c/V_b+1=60$) the smallest measurable pressure change is \SI{0.024}{Pa}, while the largest positive and negative measurable pressure changes are \SI{1400}{Pa} and \SI{-1.9e4}{Pa}, respectively. Therefore by only changing the additional volume $V_c$, the pressure measurement range can be tuned in a huge range while preserving outstanding dynamic range of $\sim10^5$ (Fig. \ref{fig9}f).

	These values far exceed the performance of the best pressure sensors of a comparable size \cite{sahota2020fiber,vorathin2020review,domingues2018cost,zhu2013graphene,wang2016graphene,javed2019review,lin2018ultra,vsivskins2020sensitive,chen2019ultrahigh} (Supplementary Table S1). Especially the dynamic range of the bubble laser pressure sensors is several orders of magnitude larger than other pressure microsensors demonstrated till now. Further, the significant advantage of the bubbles over pressure sensors using solid transducers, such as strain gauges and diaphragms, is that the smectic layer is very soft as well as practically infinitely stretchable, due to its fluid nature. The bubbles also do not experience any material fatigue. Further, the bubble pressure sensors do not need any calibration, since their sensitivity depends only on the compressibility of the air (for $V_c=0$). This is in stark contrast to almost all sensors, where variability in the mechanical properties and nonlinearities of the material (e.g. diaphragms), hysteresis and manufacturing tolerances play a crucial role in the accuracy of these sensors.

	\section*{Discussion}
	
	In conclusion, we demonstrated lasing in soap and smectic bubbles. Compared to solid or droplet microcavities, bubbles are made of a very thin film of fluid, resulting in truly unique optical and mechanical properties. The size of a bubble laser can easily be changed by a factor of 10 or more in real time. Only a multimode fiber and a simple lens are needed to pump the bubble and observe the lasing. The smectic bubbles have a completely uniform thickness down to a molecular level. The thin walls support only a single optical mode, therefore they can be employed as tunable laser sources with frequency-comb like emission, with orders of magnitude larger tunability compared to solid microlasers. Small thickness also results in small weight and therefore almost perfectly spherical shape. The bubbles are extremely stable in time and could in principle survive indefinitely. Lastly, the bubbles are very easy to form, not requiring any complicated manufacturing procedures and if a bubble is destroyed, a new one can be formed within seconds. These outstanding properties enable a number of prominent applications. Their exceptional sensitivity enables sensing of electric field and pressure with record high sensitivity, resolution and dynamic range. In future, they could be also used to measure other quantities, which change the bubble shape, such as air flow or magnetic field by doping with magnetic particles. On the opposite, to enhance the stability of the bubbles, they could be made solid, by either cooling them to crystallize or using polymerizable liquid crystals or liquid crystal elastomers \cite{jampani2019liquid}. Apart from the demonstrated applications, the bubble lasers could be used in future to study thin films and basic phenomena such as for example cavity optomechanics \cite{aspelmeyer2014cavity,kim2013cavity,asano2016observation}.

	\subsection*{Materials and Methods}
	
	A mixture of water, glycerol and liquid hand soap (containing sodium laureth sulfate) in a 2:1:1 volume ratio was used to make the soap bubbles ($n=1.364$). Alternatively, 2.5\% of sodium dodecyl sulfate dissolved in 1:1 volume ratio of water and glycerol was used. For the gain, 0.1\% of fluorescein sodium salt was dissolved in the above mixtures. For the smectic bubbles, 4'-octyl-4-biphenylcarbonitrile (8CB) liquid crystal doped with 0.2\% Pyrromethene 597 (Exiton) was used. Plastic pipette tips (Eppendorf) of different sizes (\SI{10}{\micro l}, \SI{20}{\micro l} and \SI{100}{\micro l}) were used as capillaries to inflate the bubbles. The pipette tip was connected by a thin tube to a glass syringe (Hamilton, \SI{1.0}{ml}). The syringe was mounted onto a microfluidic syringe pump (New Era Pump Systems, NE-1002X) or pushed by hand to inflate the bubble. The syringe and the connection tube to the pipette tip were filled with water, except for the last few millimeters in order to decrease the air volume. The capillary was dipped into the solution so that a small amount entered the capillary. For electric field tuning the bubble was placed inbetween two flat electrodes with an area of $18 \times$\SI{20}{mm} and a spacing of \SI{5.46}{mm}. One of the electrodes had a hole with a diameter of \SI{2.6}{mm} through which the pipette tip was inserted so that the bubble was in the center between the electrodes. For pressure measurements the pipette tip was inserted through a hole into a transparent container, which was connected to a pressure controller (Elveflow, OB1, \SIrange{0}{20}{kPa}). The bubbles were observed with an inverted microscope (Nikon Ti2) through a $4\times$, \SI{0.13}{NA} objective. The bubbles were pumped with a nanosecond pulsed optical parametric oscillator (Opotek, Opolette 355) at \SI{494}{nm} for the soup bubbles and \SI{525}{nm} for the smectic bubbles, at a repetition rate of \SI{20}{Hz}. The resulting fluorescent light was captured by a high resolution spectrometer (Andor Shamrock SR-500i) at \SI{0.007}{nm} spectral resolution.

	\section*{Author Contributions}
	Z.K. performed the experiments. Z.K. and M.H. analyzed the data and wrote the manuscript. M.H. designed and supervised the study.
	
	\section*{Acknowledgments}
	This project has received funding from the European Research Council (ERC) under the European Union’s Horizon 2020 research and innovation programme (grant agreement No. 851143) and from Slovenian Research Agency (ARRS) (J1-1697 and P1-0099).
	
	\section*{Competing interests}
	The authors declare no competing interests.
	
	\section*{Data availability}
	All data that support the plots within this paper and other findings of this study are available from the corresponding author upon reasonable request.
	
	\section*{Code availability}
	The codes that support the findings of this study are available from the corresponding author upon reasonable request.

	\section*{APPENDIX A: Air permeation through smectic bubbles}
	The permeation of gas through the smectic film is described in \cite{li2002gas}. It depends on a phenomenological parameter of gas permeation $\gamma$ describing the volume of gas passing through the film per pressure difference and area. The equation that describes the radius time dependence is 
 \begin{equation}
	R^2(t) = R_0^2 - gt, \label{R(t)_approx}
	\end{equation}
	with the initial radius $R_0=R(0)$ and the fitting parameter $g(d,T)=8\sigma\gamma $. For longer times, the first-order correction takes into account  that the area of the bubble is smaller due to the capillary opening ($r_c$). This results in a corrected equation for the size time dependence \cite{li2002gas}:
	\begin{equation}
	R^2+\frac{r_c^2}{4}\ln{\left(R^2-\frac{r_c^2}{4}\right)}=R_0^2+\frac{r_c^2}{4}\ln{\left(R_0^2-\frac{r_c^2}{4}\right)}-gt.
	\label{R(t)}
	\end{equation}
	
	The experimental data of the smectic bubble radius shrinkage over time were fitted with Eq. \eqref{R(t)}.  The fitted parameters are $R_0=$ \SI{592}{\um}, $r_c=$ \SI{336}{\um} and $g=3.85\times 10^{-10} \mathrm{m^2/s}$. The gas permeation parameter $\gamma$ obtained from the fit is $\gamma=(2.0\pm0.1)\times 10^{-9} \mathrm{m/Pa\,s}$, where the surface tension value for 8CB smectic is assumed to be $\sigma=$ \SI{0.024}{N/m} \cite{stannarius1997surface}.
	The good agreement of data and fit indicates that there was no leakage of air through the apparatus for bubble inflation, since the time dependence of the radius would be different in the case of leakage ($R^4 - R_0^4 \propto t$) \cite{li2002gas}.

	The permeation of gas through smectic film depends strongly on the film thickness and temperature, as well as on the species of the gas molecules \cite{ishii2009gas, li2002gas}. The permeation of gas molecules follows Fick's law, which describes the flow of a gaseous solute within a liquid solvent \cite{ishii2009gas}:
	\begin{equation}
	J=P\,\frac{\Delta p}{d},
	\end{equation}
	where $P=DS$ is the permeability coefficient of the gas through the smectic film and depends on the diffusion constant and the solubility of the gas molecules in the smectic, $\Delta p$ is the pressure difference between the sides of the film, and $d$ is the film thickness.
 
	If we combine the measured value of $\gamma$ with the permeability coefficient $P$ from the model, we can write:
	\begin{equation}
	\gamma=P/d.
	\end{equation}
	Using the known permeability coefficients $P$ for different gases from the study \cite{ishii2009gas}, we can calculate the thickness of the observed bubble.
	Assuming that air is composed of 78\% nitrogen and 22\% oxygen, we use the permeability coefficients $P_{\mathrm{N_2}}=(3.8\pm0.3)\times 10^{-15}\,\mathrm{mol\,Pa^{-1}m^{-1}s^{-1}}$ and $P_{\mathrm{O_2}}=(1.1\pm0.1)\times 10^{-14}\,\mathrm{mol\,Pa^{-1}m^{-1}s^{-1}}$ to calculate the effective permeability coefficient for air $P_{\mathrm{air}}=(5.3\pm0.5)\times 10^{-15}\mathrm{mol\,Pa^{-1}m^{-1}s^{-1}}$.
	Expressing $\gamma$ in units of moles instead of the volume of air exiting the bubble, we can calculate the thickness of the bubble as
	\begin{equation}
	d=\frac{(5.3\pm0.5)\times 10^{-15}\,\mathrm{mol\,Pa^{-1}m^{-1}s^{-1}}}{(8.1\pm0.4)\times 10^{-8}\,\mathrm{mol\,Pa^{-1}m^{-2}s^{-1}}}= \SI{65\pm9}{nm}.
	\end{equation}
	This value is consistent with the value calculated from the measured effective refractive index.

	\section*{APPENDIX B: Temperature change effects}
Since we assume that the size of the bubble changes according to the ideal gas law
\begin{equation}
    \frac{V'}{V}=\frac{p_iT'}{p_i'T},
\end{equation}
the temperature change ($T'-T$) of the air inside the bubble has the same effect on the bubble volume as the pressure change ($p_i'-p_i$).
The spectral shift for both pressure and temperature change is $\Delta \lambda/\lambda \approx - ( \Delta p/ 3 p_0 - \Delta T/ 3 T_0)$, where $\Delta p$ is the pressure change, $p_0$ is the initial pressure, $\Delta T$ is the temperature change and $T_0$ is the initial temperature.
The equation \eqref{dp} that relates the change in external pressure to the change in bubble size taking into account the temperature change is 
\begin{equation}
	\Delta p = \frac{V}{V'}{\left(1+\frac{\Delta T}{T}\right)}p_L - p_L' + p_0\left[\frac{V}{V'}{\left(1+\frac{\Delta T}{T}\right)}-1\right]. 
	\end{equation}
 In the approximation of small radius change, the relation between the small size, pressure and temperature change is
 \begin{equation}
	\Delta R= R\left[\frac{3\,(1+\Delta T/T)}{1+V_c/V_b}\,\left(p_0+\frac{4\sigma}{R}\right) - \frac{4\sigma}{R}\right]^{-1}\left[\frac{\Delta T}{T}\left(p_0+\frac{4\sigma}{R}\right)-\Delta p\right].
	\end{equation}
 
Apart from the temperatures changes in the environment, the pump laser could have an effect as well. To estimate the temperature rise of the air in the bubble we consider the total absorbed pump laser energy which depends on the concentration of the dye in the bubble and bubble thickness.
The extinction coefficient of fluorescent dye in smectic bubbles is \SI{80000}{M^{-1}cm^{-1}} \cite{banuelos2004photophysical} and its concentration is \SI{5.3}{mM}.
Since the bubbles are very thin (\SI{\sim50}{nm}) only a small fraction of the incident pump energy is absorbed ($\sim$0.2\%).
A fraction of that energy is converted into fluorescence light so the remaining energy that is converted into heat is given by $Q=(1-\mathrm{QY})E_{abs}$, with the quantum yield (QY) of the fluorescent dye is 0.77 \cite{vellaichamy2023optical}.
At pump pulse energy of \SI{3}{\micro J} that gives us the total heat energy of \SI{2}{nJ}.
We assume that this energy is evenly distributed across the whole bubble with the heat capacity of smectic layer \SI{1.3}{\micro J/K} and the air inside (heat capacity \SI{5.1}{\micro J/K}). This gives a temperature increase of $\Delta T=$ \SI{0.3}{mK} for each laser pulse, which is a relative change of \SI{e-6}{} compared to room temperature. That is lower than the relative pressure resolution (\SI{5e-6}{}), so it does not limit our measurements. Also during the experiments the pump laser was always turned on and the bubble was in the stationary regime meaning there are no changes in the temperature during the measurement. It could only have an effect as the measurement is started and the laser is turned on. In order to test this, lasing was observed immediately after switching on the pump laser (Supplementary Fig. S6). Apart from the change in size due to air diffusion, there was no additional size change due to the increase in temperature on a short timescale. Even with longer observations the increase in temperature seems to have no effect (Supplementary Fig. S5).

\clearpage

\setcounter{figure}{0}

\section*{Supplementary Information}

\begin{figure}[h]
	\begin{center}
		\includegraphics[width=10cm]{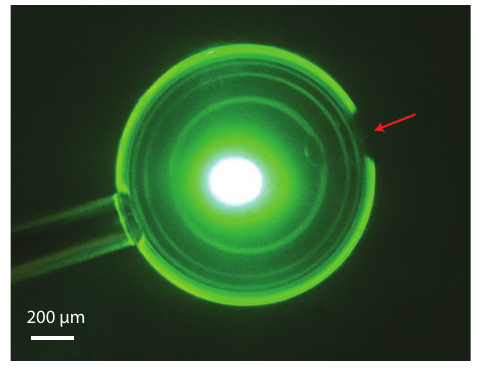}
		\caption{A soap bubble pumped at the center. Light only circulates in planes intersecting with the pump spot, but not intersecting the capillary, which can be seen as a dark patch on the opposite side of the capillary (red arrow).}
		\label{fig1}
	\end{center}
\end{figure}

\clearpage

\begin{figure}[h]
	\begin{center}
		\includegraphics[width=13cm]{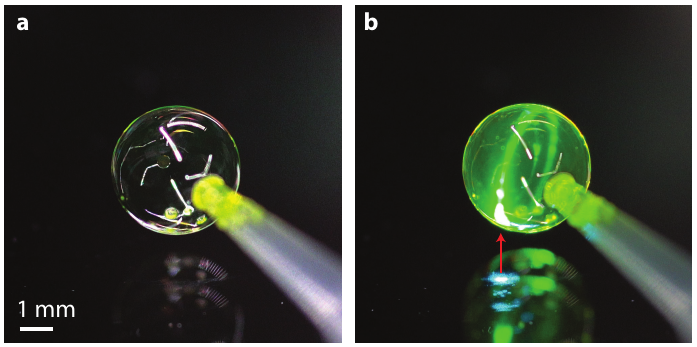}
		\caption{A dye doped soap bubble at the end of a capillary. (a) In white light interference colors can be observed. (b) The same bubble when illuminated by the pump laser. The arrow shows the direction of the incoming laser beam. A bright ring, which crosses the laser spot, is formed due to circulating light.}
		\label{fig2}
	\end{center}
\end{figure}

\clearpage

\begin{figure}[h]
	\begin{center}
		\includegraphics[width=15cm]{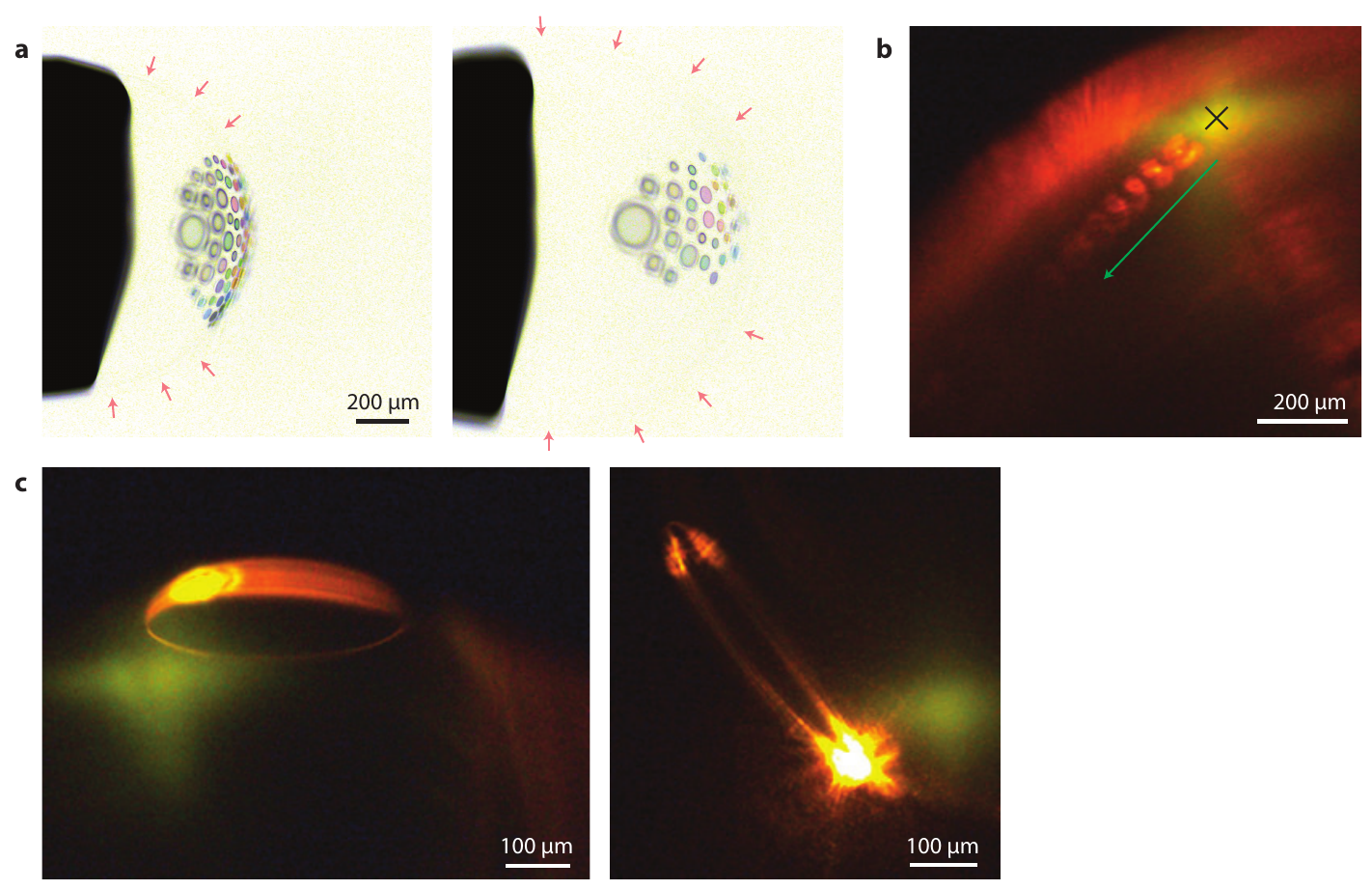}
		\caption{Smectic island lasers. (a) Two examples of smectic islands (regions of higher thickness) floating on a smectic bubble. Each island has a different thickness, therefore each displays a different interference color. The smectic bubble (marked by the red arrows) is very thin and despite the enhanced contrast of the image, it is barely visible. (b) The islands can also be created by illuminating a thicker bubble ($>\SI{200}{nm}$) with a pulsed laser (black cross), where each pulse can create one island. The islands move away from the laser spot (green arrow). (c) Two examples of lasing from smectic islands, which are visible as bright rings.}
		\label{fig3}
	\end{center}
\end{figure}

\clearpage

\begin{figure}[h]
	\begin{center}
		\includegraphics[width=10cm]{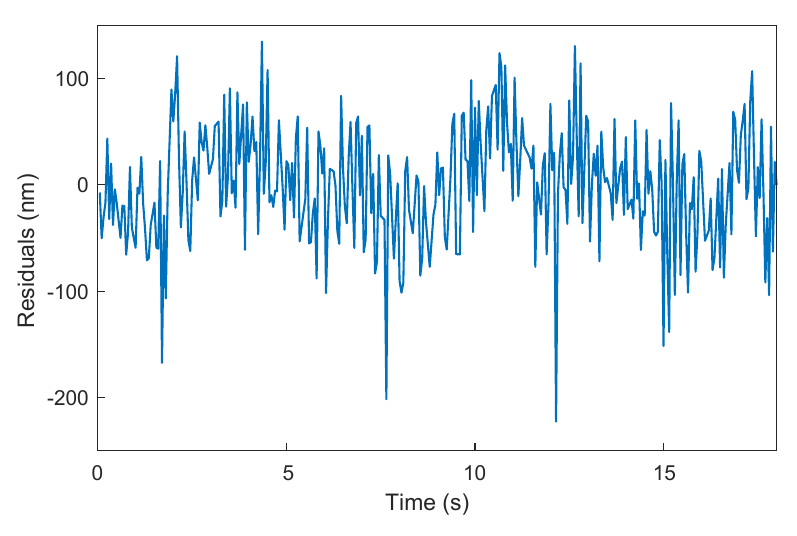}
		\caption{Data from the same bubble as in main text Fig. 7b and c. The difference here is that size of the smectic bubble was calculated from the FSR, not by following shifts of modes. The decrease in the size due to air diffusion was subtracted from the data. The remaining residuals have a standard deviation of \SI{50}{nm}.}
		\label{fig4}
	\end{center}
\end{figure}

\clearpage

\begin{figure}[h]
	\begin{center}
		\includegraphics[width=10cm]{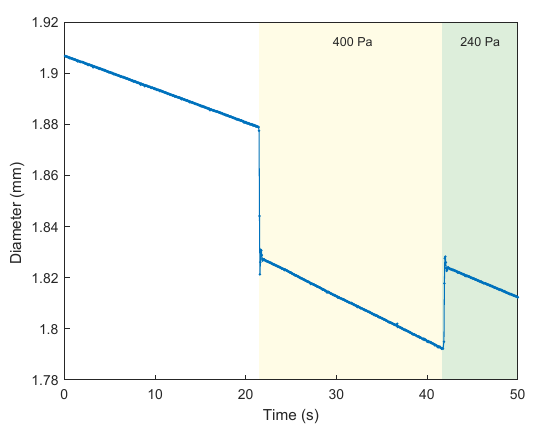}
		\caption{Diameter of a bubble in time measured from FSR. The pressure outside the bubble is increased to \SI{400}{Pa} above the atmospheric pressure and then decreased to \SI{240}{Pa}. The continuous decrease in size is due to air permeation through the smectic layer.}
		\label{fig5}
	\end{center}
\end{figure}

\clearpage
%\begin{landscape}

\begin{table}[h]
	\begin{center}
		\begin{tabular}{p{6cm}@{\hspace{1cm}} p{2cm} p{2cm} p{2cm} p{2cm} }
			\hline
			Pressure sensor & Resolution (Pa) & Maximum pressure (kPa) & Dynamic range & Size\\ [20pt] 
			\hline\hline
			Smectic bubble laser ($V_c=0$) - this work & 1.5 & 10000 & $6.7\times 10^6$&\SI{2}{mm}  \\ [20pt]
			\hline
			Smectic bubble laser ($V_c/V_b+1=60$) - this work & 0.024 & 20 & $8.3\times 10^5$ & \SI{2}{mm}\\ [20pt]
			\hline
			Percolative metal nanoparticle arrays \cite{chen2019ultrahigh} & 0.5 & 30 & $6\times 10^4$ & 1-\SI{10}{mm} \\ [20pt]
			\hline
			Temperature-insensitive optical fiber pressure sensor \cite{xu2005novel} & 69 & 1380 & $2\times 10^4$ & \SI{1.5}{mm} \\ [20pt]
			\hline
			Rocking filters in microstructured fibers \cite{anuszkiewicz2012sensing} & 300 & 10000 & $3.3\times 10^3$ & \SI{16}{cm} \\ [20pt]
			\hline
			Micro electro mechanical system sensor \cite{takahashi2010differential} & 0.1 & $\pm 0.1$ & $1\times 10^3$ & \SI{1.5}{mm} \\ [20pt]
			\hline
			Rubber diaphragm fiber Bragg grating pressure transducer \cite{allwood2015highly} & 50 & 15 & $3\times 10^2$ & \SI{14}{mm} \\ [20pt]
			\hline
			Fiber-optic Fabry–Perot interferometric pressure sensor \cite{zhang2013high} & 180 & 690 & $2.6\times 10^2$ & \SI{100}{\micro m} \\ [20pt] 
			\hline
			Graphene based piezoresistive pressure sensor \cite{zhu2013graphene} & 1500 & 70 & $4.7\times 10^1$ & \SI{280}{\micro m} \\ [20pt]
			\hline
		\end{tabular}
	\end{center}
	\caption{The comparison of smectic bubble laser as pressure sensor to different pressure sensors of comparable size.}
	\label{tab1}
\end{table}

%\end{landscape}

\clearpage

\begin{figure}[h]
	\begin{center}
		\includegraphics[width=11cm]{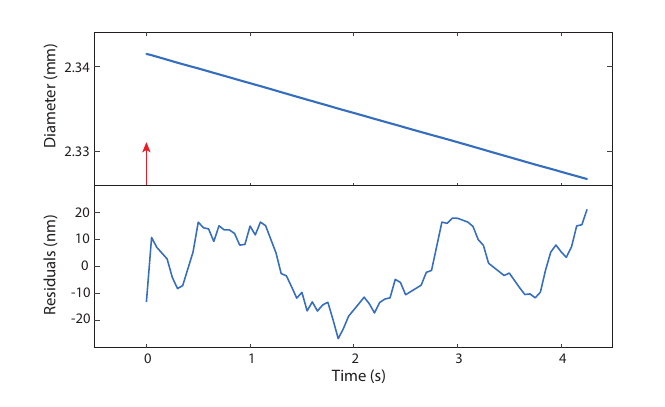}
		\caption{Change in the diameter of a bubble in time measured by following the lasing peaks immediately after switching on the pump laser (red arrow). The bubble size was slowly decreasing due to air diffusion through the thin wall but there was no additional change in size due to the increase in temperature. When the decrease in size due to air diffusion is subtracted from the data, only a small noise remains and have no larger deviations right after switching on the pump laser.}
		\label{fig10}
	\end{center}
\end{figure}
\clearpage

\begin{figure}[]
	\renewcommand{\figurename}{Supplementary Video 1}
	\renewcommand{\thefigure}{\hspace{-.333333em}}
	\centering
	\caption{A soap bubble is inflated at the tip of a capillary. At first the bubble is illuminated by a LED to see the fluorescence. The LED illumination then is turned off and the pump laser is turned on. The soap bubble starts emitting laser light. The bubble is moved so that the pump beam position is changed. The circulating laser light due to WGMs is observed as bright rings, which change their position depending on the location of the pump laser on the bubble.}
\end{figure}

\begin{figure}[h]
	\renewcommand{\figurename}{Supplementary Video 2}
	\renewcommand{\thefigure}{\hspace{-.333333em}}
	\centering
	\caption{A soap bubble is blown into a tank filled with CO$_2$. The bubble floats on the CO$_2$ and slowly changes in color due to thinning of the soap film. After a few seconds the bubble bursts. Lasing of three freely floating dye doped soap bubbles is shown when they are pumped with an external laser via an optical fiber and a lens.}
\end{figure}

\begin{figure}[h]
	\renewcommand{\figurename}{Supplementary Video 3}
	\renewcommand{\thefigure}{\hspace{-.333333em}}
	\centering
	\caption{Islands of larger thicknesses floating on a smectic bubble. The islands move due to the air currents.}
\end{figure}

	\bibliographystyle{unsrt}

\end{document}